\documentclass[preprint,12pt,authoryear]{elsarticle}

\usepackage{amsmath}
\usepackage{amssymb}
\usepackage{lineno}
\usepackage[inline] {trackchanges}
\usepackage{xcolor}
\journal{Journal of Hydrology}

\begin{document}

\begin{frontmatter}

\title{Synthetic Seismograms from Particle–Bed Interactions and Turbulent River Flow: Modeling and Comparison with Observations}

\author[inst1]{Sara Nicoletti}

\affiliation[inst1]{organization={University of Florence, Department of Mathematics and Computer Science},
   addressline={Viale Giovanni Battista Morgagni, 67/a}, 
   city={Florence},
   postcode={50134}, 
   state={Italy}}

\author[inst2]{Giacomo Belli}
\author[inst1]{Omar Morandi}
\author[inst2]{Emanuele Marchetti}

\affiliation[inst2]{organization={University of Florence, Department of Earth Science},
   addressline={via G. La Pira, 4}, 
   city={Florence},
   postcode={50121}, 
   state={Italy}}

\begin{abstract}

We present a physics-based numerical model that estimates the seismic radiation generated by water–sediment flows in gravel-bed rivers. The model reproduces the trajectories of individual particles, evaluates impact and rolling forces from grain-scale dynamics, and  accounts for broadband turbulence and vortex shedding in the water column. Synthetic seismic signals are propagated to the receivers using the Rayleigh-wave Green’s function approach and synthetic ground-velocity signals are estimated. Application to a controlled test case shows how intermittent, size-selective sediment transport mechanisms produce distinct spectral signatures. Comparison with seismic data from a flood event in a mountain torrent in the Tuscan Apennines displays general agreement with the observed frequency bands and clarifies the relative width of particle collisions and turbulent flow. These results show that resolved grain-scale dynamics provides a framework for discriminating  sediment-transport and flow-induced contributions to river seismic noise.
\end{abstract}

\begin{highlights}
\item A physics-based model computes river seismic signals from bedload and water.
\item The model reproduces frequency signatures of size-selective sediment transport.
\item Synthetic records show good agreement with field data from a mountain torrent.
\end{highlights}

\begin{keyword}
Fluvial seismology \sep Bedload transport \sep Turbulent river flow \sep Sediment transport modeling \sep Synthetic seismogram 
\sep Particle-bed interactions
\end{keyword}

\end{frontmatter}


\section{Introduction}

In gravel-bed rivers and mountain torrents, solid-particle transport plays a central role in sediment transfer throughout the fluvial network and in the development of channel morphology, ultimately regulating catchment-scale erosion and landscape evolution \citep{Gomez1989,Gomez1991,Bakker2020}. Understanding bedload transport is therefore essential for quantifying sediment fluxes, interpreting river dynamics, and supporting engineering applications \citep{Rickenmann2017}.\\

Traditionally, bedload transport has been measured using traps or samplers deployed within the flow to capture material moving along the riverbed \citep{Bunte2004,Bergman2007,Mclaughlin2025}. However, these measurements are difficult to perform continuously, particularly during high-flow events, and generally provide limited spatial and temporal resolution \citep{Mizuyama2010,Claude2012,Rickenmann2017}. 

An alternative approach estimates sediment fluxes from particle impacts recorded by geophones or force plates embedded in the channel bed \citep{McArdell2007,Turowski2009,Rickenmann2012,Rickenmann2017,Roth2016}. Although effective, this method requires calibration against independent bedload measurements and involves significant installation and maintenance efforts, while remaining restricted to the monitored river section \citep{McArdell2007,Antoniazza2020,Rickenmann2012,Rickenmann2014}. Consequently, our knowledge of bedload transport processes, particularly in gravel-bed rivers and mountain torrents, remains incomplete \citep{Rickenmann2017}.\\

Over the last two decades, seismometers have emerged as a promising tool to monitor and study bedload transport in rivers \citep{Burtin2008,Burtin2011,Hsu2011,Tsai2012,Barriere2015b,Bakker2020}. Transported sediment particles interact mechanically with the riverbed, generating stress oscillations that propagate as seismic surface waves \citep{Burtin2008,tsai2009,Tsai2012,Bakker2020,Belli2025}. These signals originate mainly from grain-bed impacts and frictional contacts such as rolling or sliding \citep{Tsai2012} and can be used to infer bedload fluxes and transport dynamics from seismic recordings acquired near the channel \citep{Schmandt2017,Bakker2020,Johannot2025,Belli2026}.\\

However, particle impacts are not the only seismic source active in rivers. Flow waves and turbulent fluctuations also radiate seismic waves into the ground \citep{Ervine1997,Schmandt2013,Gimbert2016,Belli2025}, and recorded signals reflect contributions from both the solid and liquid phases. Turbulence generally produces lower-frequency seismic signals than sediment transport \citep{Burtin2011,Schmandt2013,Gimbert2016}. Nevertheless, this separation can be blurred by wave propagation effects \citep{Tsai2012} and may break down in small mountain streams, where finer turbulent structures lead to overlapping frequency bands between the two sources \citep{Roth2016}. Distinguishing seismic signals generated by particle impacts from those produced by water flow therefore remains challenging \citep{Bakker2020,Roth2016}.\\

Progress in addressing this problem can be achieved through the physical modeling of seismic source processes \citep{Tsai2012,Gimbert2014} and through numerical simulations that reproduce the flow dynamics responsible for seismic signal generation. A key challenge is to establish a quantitative and physically consistent link between grain-scale sediment dynamics, turbulent flow processes, and the resulting seismic radiation. Previous studies have identified characteristic frequency bands associated with bedload transport and turbulent flow (e.g., \citealp{Hsu2011,Tsai2012,Schmandt2013}), while numerical and laboratory investigations have highlighted the role of particle impacts and rolling contacts in generating high-frequency signals (e.g., \citealp{Hsu2004,Schmeeckle2014,Gimbert2014}). Many existing approaches rely on simplifying assumptions about complex natural processes \citep{Tsai2012,Gimbert2014,Coco2021}, limiting their ability to disentangle the respective contributions of particle collisions and hydrodynamic forcing. Field validation against observational data also remains limited \citep{Gimbert2016,Bakker2020,Luong2024}.\\
Water-driven seismic forcing in rivers is generally modeled incorporating both broadband pressure fluctuations associated with turbulent shear stresses and flow waves, as well as narrow-band tones generated by vortex shedding, exert dynamic forcing on the riverbed \citep{Gimbert2014,Ervine1997,Gimbert2016,Tsai2012water}. Physically based models \citep{Gimbert2014} represent these processes by combining broadband turbulence consistent with Kolmogorov scaling \citep{kolmogorov1941}, with coherent vortex-shedding contributions parameterized through Strouhal-type relations \citep{strouhal1878}.

Most existing models of seismic radiation by bedload transport adopt analytical or semi-analytical approaches, in which sediment transport is represented statistically, through prescribed impact rates or stochastic source terms \citep{Tsai2012,Gimbert2014}. Both impulsive grain impacts and sustained frictional motion have been identified and are generally reproduced as seismic sources in the flow, often producing overlapping frequency bands. Although such formulations allow one to estimate empirical scaling laws for bedload flux, grain size, and seismic power spectra, they do not attempt to reproduce the individual trajectories of sediment particles.
Particle-resolving numerical models based on Lagrangian or discrete element methods (DEM) have been developed to simulate sediment transport by explicitly computing grain dynamics under gravity, hydrodynamic drag, collisions, and frictional contacts \citep{Cundall1979,Maurin2015}. These approaches provide microscopic-scale resolution of grain-scale mechanics but have only rarely—and, to our knowledge, never in a fully resolved manner—been coupled with forward seismic modeling. As a result, they do not directly predict the seismic signals generated by particle motion.

Conversely, seismic source models for rivers commonly represent hydrodynamic forcing as a stochastic bed stress mechanism associated with turbulent pressure fluctuations, without explicitly accounting for the resolved kinematics of mobile sediment particles \citep{Gimbert2014,Schmandt2013}.

As a result, there exists a gap between time-resolved particle-scale models of bedload transport and physically based seismic source models. Bridging this gap requires developing theoretical and numerical approaches that simultaneously resolve individual particle trajectories and compute the resulting seismic signal via forward models that map particle-induced forces into ground motion, such as the transfer function formulation proposed by \citet{Tsai2011}. Such a fully resolved approach is essential to disentangle the relative contributions of sediment transport and turbulent flow to river seismic noise and to assess whether, and to what extent, bedload transport properties can be inferred from field seismic observations.

We describe our numerical model of seismic radiation generated by a water–sediment flow over a two-dimensional inclined rough bed. Based on the approaches of \citet{Tsai2012} and \citet{Gimbert2014}, the model simulates seismic signals produced by both bedload transport and water flow. The resulting synthetic seismograms are compared with seismic data recorded in correspondence of two stream sections during a flood event in a mountain torrent in the Tuscan Apennines. This comparison allows us to quantify the relative contributions of particle impacts, rolling motion, and turbulent flow forcing to the total seismic power spectral density (PSD), and to assess how grain-scale transport processes are reflected in the observed seismic spectrum. General agreement is found between simulations and observations, validating the ability of numerical model to reproduce the key features of the river seismic radiation. This framework provides a direct bridge between particle-scale sediment dynamics and river-generated seismic observations, enabling a physically consistent interpretation of seismic spectra in terms of underlying transport processes.



\section{Methods}
\label{Methods}
In this work we develop a physics-based numerical model that explicitly resolves the time-dependent trajectories of individual sediment particles and directly links grain-scale dynamics to forward seismic modeling. The motion of spherical particles is reproduced under the combined action of gravity, hydrodynamic drag, grain–bed and inter-particle collisions, frictional contacts, and stochastic forcing representing near-bed turbulent fluctuations. The simulation of the particle trajectories is resolved in time, allowing to classify dynamically the interactions with the bed into impact- and rolling-dominated regimes, without prescribing impact statistics a priori.

The forces generated by particle impacts, rolling contacts, and turbulent flow are then used as source of seismic signal, estimated by forward modeling framework. In particular, water-induced forcing is represented through broadband turbulent fluctuations which are consistent with inertial-range scaling \citep{kolmogorov1941} together with narrow-band vortex-shedding contributions. All source terms are propagated to a virtual receiver using analytical Rayleigh-wave Green’s function approach \citep{Tsai2011}, enabling the generation of synthetic seismograms.

The numerical procedure used to generate synthetic seismograms consists
of the following steps: (i) compute particle trajectories; (ii) detect
contacts with the bed and classify them as impacts or rolling events;
(iii) estimate the corresponding time-dependent contact forces exerted
on the bed; (iv) compute the hydrodynamic forcing associated with
turbulent flow and vortex shedding; (v) combine particle-induced and
water-induced forcing to obtain the total bed forcing; (vi) propagate
these forces through the Green’s function to obtain ground velocity
signals; and (vii) compute the power spectral density (PSD) of the
resulting signal, both for the total response and separately for the
individual contributions.

\subsection{Modeling the particle-induced forcing generated by particle impacts and sediment transport}
We consider spherical particles moving over a rough bed under the action of gravity, fluid drag, collisions, and stochastic forcing. The channel is initially free of sediments, and particles are injected into the domain at a prescribed rate with variable diameters. We consider a domain of length $L$ and flow depth $H$, in which particles enter the domain at a rate $N_p$ (number of particles per second). Periodic boundary conditions are imposed along the streamwise direction.

Particles are injected with zero initial velocity  and are transported by a fluid flowing with mean velocity $U_0$. Particles are injected at random vertical positions within the flow depth,
uniformly distributed between the bed and the free surface. This choice
is intended to provide a statistically homogeneous initialization of
particle positions, rather than to represent a specific physical entrainment
mechanism. The subsequent dynamics, governed by gravity, drag, and
particle–bed interactions, naturally lead to the emergence of near-bed
transport, including rolling and impact regimes.During their motion, particles may collide with the riverbed and with each other. Bed roughness is modeled by superimposing random vertical perturbations
to the bed profile, drawn from a normal distribution with zero mean and
standard deviation $\sigma_z = 1\,\mathrm{mm}$. The resulting roughness
field represents small-scale irregularities
of the bed surface. 

The position of the $i$-th particle $\mathbf{x}_i = (x_i, z_i)$ evolves according to Newton’s law

\begin{equation}
m_i \frac{d^2 \mathbf{x}_i}{dt^2} =
\mathbf{F}_{g,i}
+
\mathbf{F}_{d,i}
+
\sum_{j \neq i} \mathbf{F}_{c,ij}
+
\mathbf{F}_{b,i},
\label{eq:newton}
\end{equation}

where $\mathbf{F}_{g,i}$, $\mathbf{F}_{d,i}$, $\mathbf{F}_{c,ij}$, and $\mathbf{F}_{b,i}$ denote gravitational, drag, contact (collision), and stochastic forces, respectively.
\begin{figure}[ht!]
\centering
\includegraphics[width=0.5\linewidth]{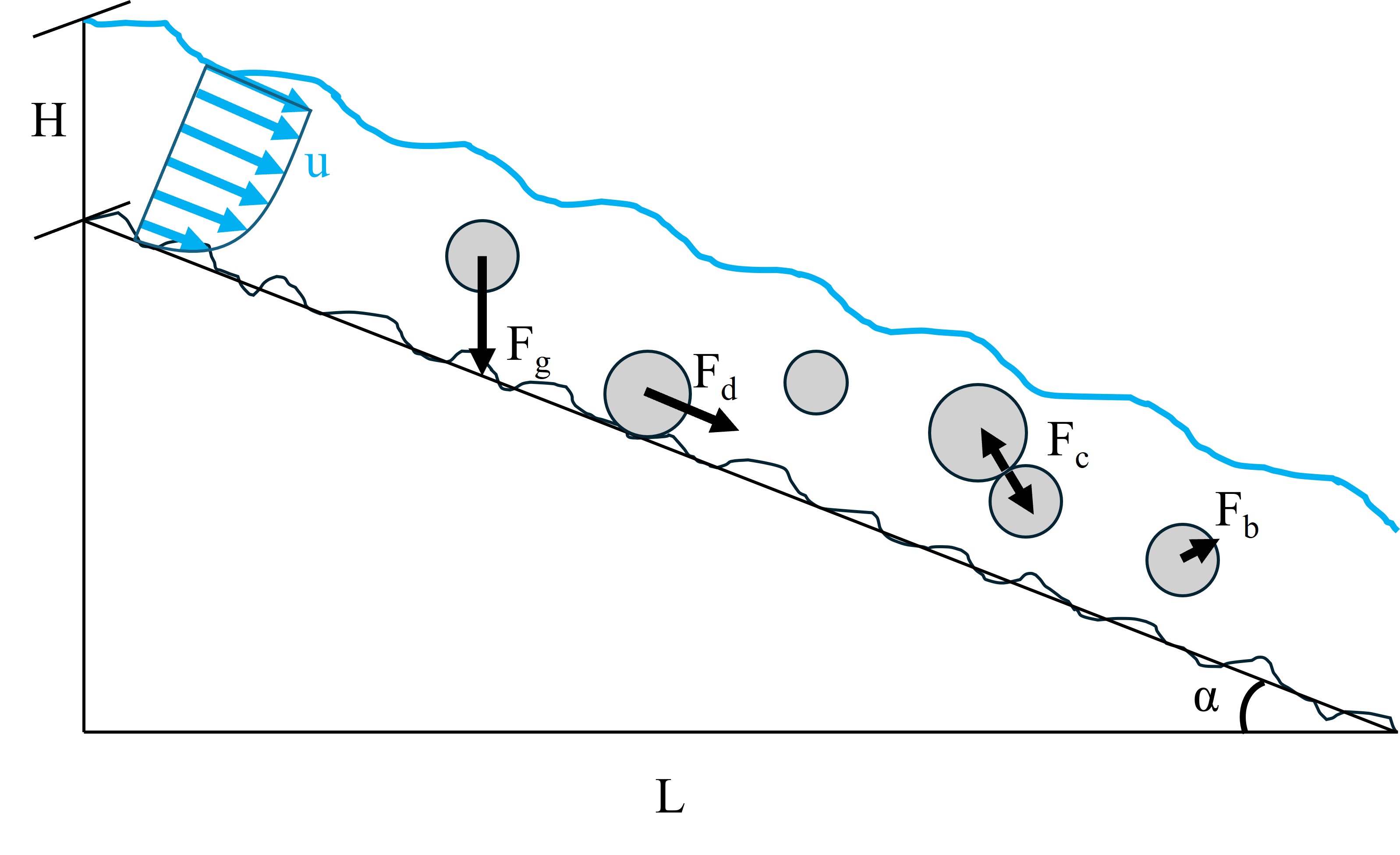}
\caption{
Two-dimensional schematic representation of the numerical model.
Spherical particles move over a rough inclined bed of slope $\alpha$
within a domain of length $L$ and flow depth $H$.
Forces acting on each particle include gravity $F_g$, hydrodynamic drag $F_d$,
inter-particle and particle–bed contact forces $F_c$, and stochastic
forcing $F_b$ representing unresolved near-bed flow fluctuations.
Bed roughness is modeled as a random vertical perturbation.
}
\label{test}
\end{figure}
The gravitational force is given by $\mathbf{F}_{g,i} = m_i \mathbf{g}$.
For spherical particles, the drag force is computed using the classical correlation of \citet{SchillerNaumann1935} $\mathbf{F}_{d,i}
=
C_D \frac{\pi D_i^2}{8}
\rho_f
|\mathbf{u}_f-\mathbf{v}_i|
(\mathbf{u}_f-\mathbf{v}_i),$ where $D_i$ is the particle diameter, $\rho_f$ the fluid density, $\mathbf{u}_f$ the local fluid velocity, and $\mathbf{v}_i$ the particle velocity. The drag coefficient depends on the particle Reynolds number $Re_p =
\frac{\rho_f D_i |\mathbf{u}_f - \mathbf{v}_i|}{\mu_f}.$ Specifically, $C_D =
\frac{24}{Re_p}
\left(1 + 0.15\,Re_p^{0.687}\right)$
for $Re_p \le 1000,$ and $C_D = 0.44$
for $Re_p > 1000.$ In the present flow conditions, particle Reynolds numbers are
consistently larger than $10^3$, so that the drag coefficient
is effectively constant ($C_D \approx 0.44$).
The stochastic forcing term is modeled using a Langevin-type formulation $\mathbf{F}_{b,i}
=
\sqrt{\frac{2\sigma}{\Delta t}}
\,\boldsymbol{\eta},$ where $\boldsymbol{\eta}$ is a vector of independent Gaussian random variables with zero mean and unit variance, and $\Delta t$ is the numerical time step. The parameter $\sigma$ controls the intensity of the stochastic forcing and is set to $\sigma = 10^{-8}\,\mathrm{N}{\mathrm{s}}^{1/2}$ \citep{Uhlenbeck1930,Kloeden1992,Ermak1978,MaxeyRiley1983,Hoomans1996,Hoomans1996bis}. This term describes an effective stochastic forcing accounting for unresolved small-scale near-bed flow fluctuations.

The water velocity profile is prescribed. We use a  power-law model for turbulent boundary layers $U(z) =
U_{\mathrm{ref}}
\left(
\frac{z+z_0}{z_{\mathrm{ref}}+z_0}
\right)^{\alpha}.$
Here $z_0$ represents an effective hydrodynamic roughness length associated to the grain size. Following standard hydraulic approximations, we set $z_0 \approx \frac{D_{50}}{30},$ where $D_{50}$ is the median grain diameter of the bed material \citep{vanRijn1984,Soulsby1997,Julien2010}. Particle diameters are sampled from a truncated lognormal distribution
with prescribed mode $D_0 = 5\,\mathrm{cm}$ and logarithmic standard
deviation $\sigma_{\log} = 0.9$. This corresponds to a median grain
size $D_{50} \approx 11\,\mathrm{cm}$. The relatively large difference between mode and median reflects the
strong asymmetry of the grain-size distribution. The exponent $\alpha$ controls the vertical shear of the velocity profile. For fully developed turbulent boundary layers, theoretical and experimental studies show that $\alpha$ typically lies in the range $0.1$--$0.3$, and $\alpha = 1/7$ is commonly used as a practical approximation which provides a realistic representation of the shear distribution across the flow depth. \citep{Schlichting2017,Pope2000}. Accordingly, we set $\alpha = 1/7$.

\subsection*{Collisions}

Pairwise collisions between particles are modeled using a soft-sphere Discrete Element Method (DEM) formulation, in which contacts are resolved
through linear spring–damper interactions acting along the line connecting the
particle centers. Particles are treated as slightly deformable spheres that
may overlap during contact. This overlap generates an elastic restoring
force and a dissipative damping force, which are explicitly integrated in
time.

The contact model consists of normal and tangential interactions. In the
normal direction, elastic deformation and viscous dissipation are described by a linear spring and a dashpot. In the tangential direction, a
visco-elastic interaction combined with a Coulomb friction limiter accounts
for sliding resistance and energy loss.

\subsubsection*{Soft-sphere contact model}

Particle–particle and particle–bed contacts are detected by a geometric
overlap criterion. When two particles $i$ and $j$ come into contact, the
normal unit vector $\mathbf{n}$ is defined along the line connecting their
centers and the normal overlap $\delta_n$ is computed.
The normal contact force is given by $\mathbf{F}_{n,ij}
=
\left(k_n \delta_n - c_n v_n\right)\mathbf{n},$
where $k_n$ is the normal stiffness, $c_n$ the damping coefficient, and
$v_n$ the normal component of the relative velocity.
Tangential interactions are modeled by $\mathbf{F}_{t,ij}
=
- (k_t + \eta_t) v_t \mathbf{t},$
where $k_t$ and $\eta_t$ are tangential stiffness and damping coefficients,
$v_t$ is the tangential relative velocity, and $\mathbf{t}$ is the tangential
unit vector at the contact. The tangential force is limited by a Coulomb
criterion $|\mathbf{F}_{t,ij}| \le \mu |\mathbf{F}_{n,ij}|,$ where $\mu$ is the friction coefficient.
This results with the total contact force $\mathbf{F}_{ij} = \mathbf{F}_{n,ij} + \mathbf{F}_{t,ij}$.
The linear spring–dashpot formulation
captures the essential mechanisms of elastic deformation, viscous
dissipation, and frictional sliding governing granular collisions, while
remaining computationally efficient for large particle systems. 
For more refined models, we refer to Hertzian or history-dependent contact models
\citep{Cundall1979,Mindlin1953,Maw1976}. 

\subsubsection*{Particle–bed interactions}

Collisions between particles and the riverbed are modeled by an event-based
formulation, consistent with collision-based bedload and saltation models. Under such an approximation, 
particle–bed interactions are represented as short-duration momentum
exchanges triggered when specific kinematic criteria are satisfied.

Sediment--riverbed contacts are detected based on a purely geometric criterion. 
Specifically, a particle is considered in contact with the bed whenever its vertical position satisfies $z_i \leq z_{\mathrm{bed}}(x_i),$
where $z_{\mathrm{bed}}(x)$ denotes the local bed elevation, obtained via interpolation of the discrete bed profile. The interpolation is performed on a refined grid to ensure a smooth and stable evaluation of the bed elevation at particle positions.

\subsubsection*{Impact and rolling regimes}

Particle–bed interactions are dynamically classified into impact or
rolling based on particle velocities and proximity to the bed.
We assume rolling if the horizontal velocity of the sediment exceeds the
threshold $v_{x,\mathrm{roll}} \approx 0.1\,U_0$. According to experimental observations, under such a condition, the particle remains close to the bed
$(z < 1.3D)$
\citep{Lajeunesse2010,Celik2010,Diplas2008,DrakeShaw1982,Charru2013}.

Impacts are detected when the pre-collision vertical velocity exceeds
a threshold estimated as the velocity of a particle falling over one diameter altitude, $v_{z,\mathrm{impact}} \approx \beta \sqrt{2gD} + \gamma U_0 ,$
with $\beta = 0.5$ and $\gamma=0.1$ in the present simulations
\citep{Bagnold1941,Nino1998,Sun2016}.
During each impact, the vertical velocity is updated using a restitution law $v'_{z,i} = -e\,v_{z,i},$
with restitution coefficient $e = 0.45$. The associated impulse transmitted
to the bed is $J_i = m_i (1+e)|v_{z,i}^{-}|.$
Rolling contact forces are represented as weaker impulses $J_j = \alpha\,m_j(1+e)|v_{z,j}^{-}|,$
with $\alpha = 0.3$ representing reduced momentum exchange.

In the theoretical formulation of \citet{Tsai2012} and \citet{Gimbert2014}, particle–bed
collisions are represented by impulsive forces $F_i(t) = J_i \delta(t-t_i).$
In our model, we assign a finite characteristic time 
duration to each event, to account for the finite stiffness of particle–bed
contacts and to avoid unrealistically broadband excitation.
The characteristic contact time is estimated using a spring–mass
approximation $t_c = \pi \sqrt{\frac{m_{\mathrm{eff}}}{k}},$
which corresponds to half the natural period of a linear oscillator. 
%
Each event is described by a scaled Ricker wavelet
$\psi(t) =
(1-2\pi^2 f_0^2 t^2)\exp(-\pi^2 f_0^2 t^2),$
where the central frequency is obtained as the inverse of the contact time $f_0 = \frac{1}{t_c}.$
The resulting force is $F_i(t) = F_{0,i}\psi(t),$
where the amplitude $F_{0,i}$ is scaled to be consistent with the impulse $J_i$.

Rolling contacts produce low characteristic frequencies. In the numerical implementation, the contact duration associated with rolling events is artificially increased by a constant factor (here set to $10$) in order to enforce a clear separation of timescales between impulsive (impact) and sustained (rolling) interactions. This procedure does not aim at resolving the detailed contact mechanics, but rather at representing their distinct spectral signatures. Sensitivity tests show that moderate variations of this factor produce negligible changes in the resulting power spectral density.

The total particle-induced forcing is given by $F_{\mathrm{particles}}(t)
=
F_{\mathrm{impact}}(t)
+
F_{\mathrm{rolling}}(t),$.
Each component of the total forcing is obtained as the superposition of the individual force pulses generated by the corresponding contact events.



\subsection{Modeling the water forcing generated by turbulent water flow}
As indicated previously, hydrodynamic forcing is represented through prescribed velocity flow and stochastic fluctuations are consistent with turbulent boundary-layer theory.
In the numerical implementation, the seismic forcing induced by the water flow is modeled as an effective hydrodynamic load acting on the riverbed. This load (denoted by \(F_{\mathrm{water}}(t)\)) represents the fluctuating pressure and shear stresses generated by near-bed turbulence and coherent flow instabilities.

We decompose \(F_{\mathrm{water}}(t)\) into the sum of two contributions: 
(i) a broadband component associated with turbulent fluctuations and 
(ii) a narrow-band component associated with vortex shedding. 
To account for the dependence of hydrodynamic forcing on discharge, both contributions are modulated by the instantaneous flow strength.

We assume $F_{\mathrm{water}}(t)
=
\big[ U_{\mathrm{eff}}(t) \big]^p
\Big(
w_{\mathrm{turb}}\, s_{\mathrm{turb}}(t)
+
w_{\mathrm{tone}}\, s_{\mathrm{tone}}(t)
\Big),$
where \(U_{\mathrm{eff}}(t)\) is a normalized measure of the near-bed flow velocity sampled at the seismic rate (in the implementation, the spatially averaged near-bed velocity), and \(p\) is the velocity-scaling exponent. In the present work we adopt \(p=3\), consistent with energetic arguments linking turbulent power to flow velocity \citep{Gimbert2014}. The signals \(s_{\mathrm{turb}}(t)\) and \(s_{\mathrm{tone}}(t)\) are dimensionless, zero-mean, unit-RMS processes representing the temporal structure of turbulent and shedding-related forcing, respectively. The coefficients \(w_{\mathrm{turb}}\) and \(w_{\mathrm{tone}}\) (with \(w_{\mathrm{turb}}+w_{\mathrm{tone}}=1\)) are the width of the relative energy contribution of the two mechanisms.

The turbulent contribution \(s_{\mathrm{turb}}(t)\) is implemented as a stochastic signal whose spectral content mimics inertial-range turbulence. Following Kolmogorov theory \citep{kolmogorov1941}, the turbulent kinetic energy spectrum obeys \(E(f)\propto f^{-5/3}\), corresponding to an amplitude scaling \(|S(f)|\propto f^{-5/6}\). 
In practice, a white-noise signal is generated with spectral range in the interval $30-90$ Hz, with smooth tapering outside the range $0.5-100$ Hz. Similar spectral contents have been observed in river-generated seismic noise, where turbulent pressure fluctuations transmitted through the bed produce broadband seismic signals in the $10–100$ Hz range \citep{Burtin2010,Gimbert2014}.
The resulting signal represents turbulent pressure and shear-stress fluctuations acting on the bed. This formulation follows the theoretical framework developed for river-induced seismic noise by \citet{Tsai2012} and \citet{Gimbert2014}, in which pressure patches convected at the mean flow velocity excite the substrate with an inertial-range-controlled spectrum.

In addition to broadband turbulence, the interaction between the flow and bed roughness elements can generate organized vortical structures (e.g., vortex shedding). These structures can detach periodically, generating a quasi-monochromatic hydrodynamic forcing at the vortex shedding frequency $f_s = St \, \frac{U_0}{D},$
where \(St\) is the Strouhal number (\(St\approx0.2\), consistent with observations of hydrodynamic tones in fluvial environments \citep{Tsai2012,Schmandt2013}, \(U_0\) is the reference flow velocity, and \(D\) is a characteristic roughness length scale \citep{strouhal1878}.

The shedding contribution \(s_{\mathrm{tone}}(t)\) is modeled as a Gaussian narrow-band stochastic signal centered on \(f_s\). The signal is normalized to zero mean and unit RMS. Temporal variability in shedding coherence is represented by the finite spectral width of the Gaussian envelope.

\subsection{Seismic signal}

The total bed forcing is the sum of particle-induced and hydrodynamic contributions, $F_{\mathrm{bed}}(t)
=
F_{\mathrm{particles}}(t)
+
F_{\mathrm{water}}(t).$
The forces excite elastic waves that propagate through the subsurface and are recorded as ground velocity. The vertical ground velocity $u$ is computed in the frequency domain using an analytical Rayleigh-wave Green’s function \citep{tsai2009,Tsai2011}. Particle impacts and rolling events are treated as discrete, localized sources at distances \(r_s\) from the seismic station, with spectra obtained as $\tilde{u}_{\mathrm{particles}}(f)
=
\sum_s i\,2\pi f \, \tilde{F}^{(s)}(f)\, G(f;r_s),$
where \(\tilde{F}^{(s)}(f)\) is the Fourier transform of the force related to the \(s\)-th event. Hydrodynamic forcing is represented as a spatially distributed source, with bed-parallel force density \(F_{\mathrm{water}}(x,t)\) propagated via
\(\tilde{u}_{\mathrm{water}}(f) = \sum_x i\,2\pi f \, \tilde{F}_{\mathrm{water}}(x,f)\, G(f;r(x))\).

The Green’s function reads $G(f;r) = \frac{k(f)}{8\rho_m v_c v_u} \sqrt{\frac{2}{\pi k(f) r}} \exp\Big(-\pi f \frac{r}{v_u Q}\Big),$
with \(k(f) = 2\pi f / v_c\) where 
\(\rho_m = 2650~\mathrm{kg\,m^{-3}}\) is the density of the medium, \(v_c = 1300~\mathrm{m\,s^{-1}}\) and
\(v_u = 0.73\,v_c\) are the phase and group velocities of modelled Rayleigh waves, while \(Q = 20\) is the quality factor \citep{Tsai2012}.

The synthetic ground velocity in the time domain is reconstructed by inverse Fourier transform, \(u(t) = \mathcal{F}^{-1}\{\tilde{u}(f)\}\). The total signal at the virtual sensor is $u_{\mathrm{tot}}(t) = \alpha_{roll} \, u_{roll}(t) + \alpha_{imp} \, u_{imp}(t)+ \alpha_{turb} \, u_{turb}(t)+ \alpha_{shed} \, u_{shed}(t),$
where \(u_{roll}(t)\),\(u_{imp}(t)\),\(u_{turb}(t)\) and \(u_{shed}(t)\) are the velocities generated by particle rolling, impacts and hydrodynamic turbulence and shedding, respectively. The dimensionless scaling coefficients \(\alpha_{roll}\), \(\alpha_{imp}\) and \(\alpha_{turb}\), \(\alpha_{shed}\) are used to explore the relative contributions of granular and fluid processes to the observed power spectral density (PSD), accounting for uncertainties in source efficiency, bed coupling, and site-specific propagation effects.


\section{Test case}

 To explore the sensitivity of seismic signals to key parameters such as flow velocity, particle concentration and size, impact dynamics, and the relative balance between hydrodynamic and particle-induced forcing, we discuss a preliminary test case, characterized by a simplified geometry. 

Time series and power spectral densities (PSDs) are analyzed to quantify the relative contributions of particle impacts, rolling motion, and water-induced forcing. Sediment transport metrics—including particle mobility, vertical distribution, and bulk flux—are extracted from the simulated trajectories.
The test case considers a simplified two-dimensional channel geometry, whose main parameters are summarized in Table~1. The domain has length $L$, flow depth $H = 20$~cm, and bed slope of 5\%, as depicted in Figure 2, panel (a). A constant inflow of particles is imposed with rate $N_p$, and particles are characterized by a median diameter $D_{\mathrm{med}} = 5$~cm. The simulation is run over a total duration $T = 200$~s.

\begin{table}[ht!]
	\centering
	\caption{Main parameters of the numerical test case.}
	\label{tab:test_cases}
	\begin{tabular}{lcccccc}
		\hline
		$U_0$ [m/s] & $H$ [cm] & $N_p$ & $D_\mathrm{med}$ [cm] & $T$ [s] \\
		\hline
		1.0 & 20 & 2 & 5 & 200 \\
		\hline
	\end{tabular}
\end{table}

\begin{figure}[ht!]
	\centering   
	\includegraphics[width=1\linewidth]{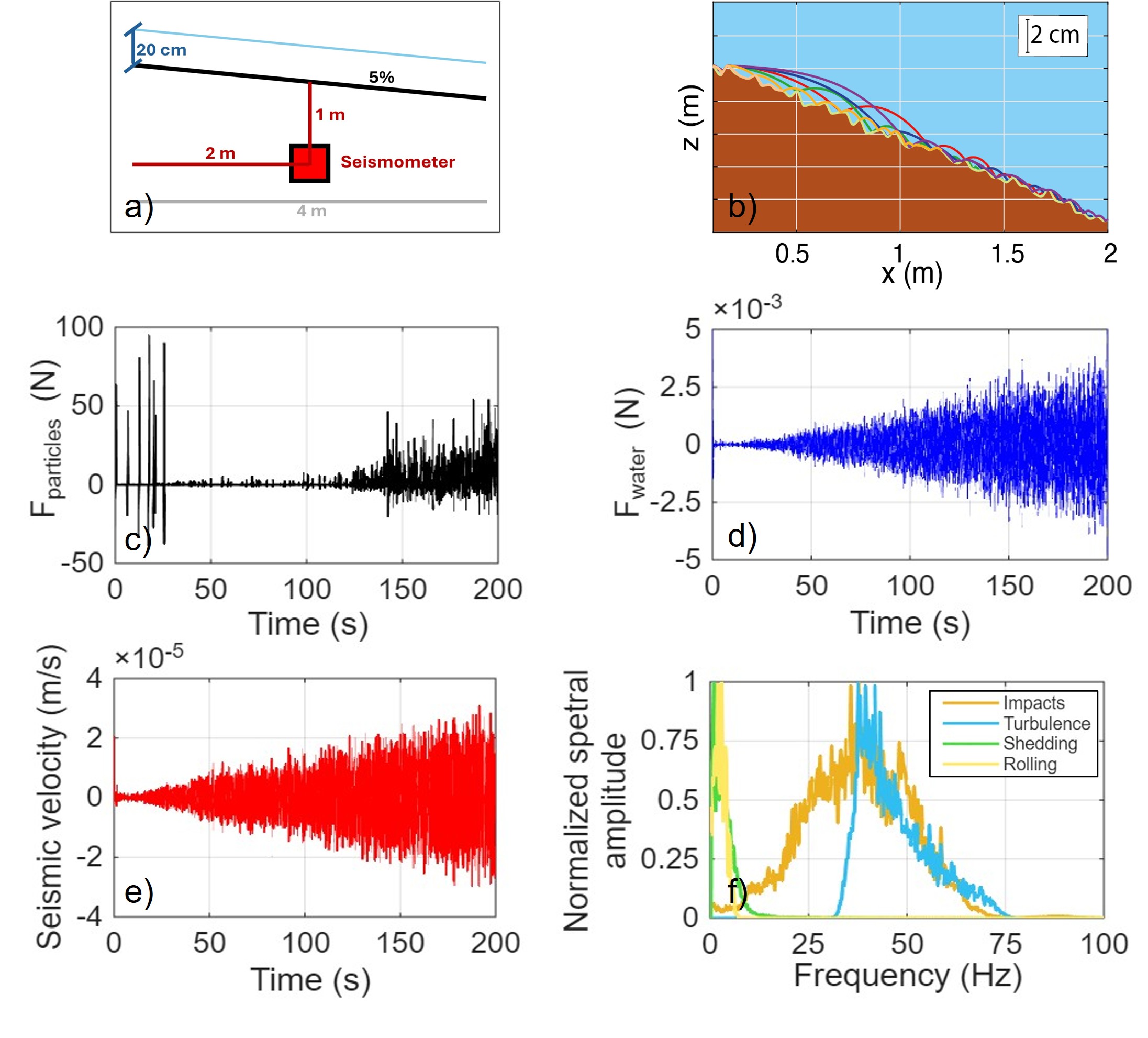}
	\caption{(a): Schematic of the numerical test domain. Flow depth is $H=20$~cm, bed slope $5\%$. The virtual seismic receiver position is indicated. (b): Simulated individual particle trajectories along the domain; the trajectories illustrate alternating phases of rolling motion along the bed  and short suspension events induced by turbulent forcing. (c): Particle-induced forcing generated by grain–bed interactions, including impulsive impacts and lower-amplitude rolling contacts. (d): Hydrodynamic forcing produced by the water flow, including broadband turbulence and vortex-shedding contributions. (e): Synthetic vertical ground velocity obtained by propagating the source forces through the analytical Green’s function. (f): Power spectral densities (PSD) of the individual source contributions, illustrating the distinct spectral signatures associated with particle impacts, rolling motion, and water-flow forcing.}
	\label{resulttest}
\end{figure}  
Figure~2, panel b, illustrates representative simulated particle trajectories within the channel during the flood conditions considered in this study. The trajectories highlight the complex dynamics of particle transport, characterized by alternating phases of rolling along the bed and short suspension events driven by turbulent flow fluctuations.
Figure~\ref{resulttest}, panels (c-d), illustrates the main forcing components. Particle-induced forces exhibit intermittent peaks associated with impacts, superimposed to lower-amplitude contributions generated by rolling contacts. In contrast, water-induced forcing appears as a more continuous broadband signal produced by turbulent flow. A temporal smoothing is applied to the forcing signal to account for the finite duration of the interaction between the turbulent flow and the bed. In practice, fluid-induced forces are not instantaneous, but persist over short time intervals, leading to a more distributed response. The smoothing reduces high-frequency fluctuations and results in a more physically consistent signal. We note that this operation may introduce a mild initial transient, as fewer data points are available for averaging at the beginning of the time series. In the absence of smoothing, the signal would be significantly noisier and dominated by sharp, nonphysical spikes.
Panels (e–f) further illustrate how these forcing components translate into seismic observations. The synthetic seismic signal (e) closely resembles the macroscopic trend of the hydrodynamic forcing (d), while the sharp impulses generated by particle impacts (visible in panel (c)) are no longer clearly distinguishable. This occurs because particle impacts have a very short contact time and generate high-frequency signals, which are strongly attenuated by the filtering effect during wave propagation, as described by the frequency-dependent exponential decay in the Green's function. Consequently, the more continuous water-induced forcing, alongside lower-frequency rolling contributions, dominates the background trend of the final seismogram, visually masking the attenuated impact transients.
The corresponding PSDs (panel f) highlight the distinct spectral signatures of the different source mechanisms. Impact-induced forcing dominates the high-frequency range (typically 10–100 Hz), while rolling motion contributes primarily at lower frequencies. Hydrodynamic forcing produces a broadband spectral component, with possible narrow-band peaks associated with vortex shedding. This spectral separation enables the identification of the underlying physical processes from the observed seismic signal.

Impact events generate short-duration force transients with characteristic contact time 
\(t_c = \pi \sqrt{m_{\mathrm{eff}}/k}\), implying dominant frequencies 
\(f_{\text{impact}} \sim 1/t_c\). For the particle masses and stiffness considered here, 
\(t_c \sim 10^{-2}\)~s, yielding frequencies in the 10--100~Hz range. 
This range is consistent with both DEM simulations \citep{Hsu2004,Schmeeckle2014} 
and field observations \citep{Tsai2012}. 

Rolling motion involves longer interaction times and smaller impulses, producing lower-frequency contributions typically below 10~Hz, consistent with advection–diffusion descriptions of bedload transport \citep{Ancey2015}.

Particles interact with the irregular channel bed and with each others, producing intermittent contacts and collisions that generate impulsive forces. These interactions are an important source of seismic radiation. Sediment transport proceeds stepwise, periods of rest or slow rolling along the bed are interrupted by entrainment into the flow, reflecting the stochastic nature of sediment transport under turbulent conditions.

Sediment transport metrics are extracted from particle trajectories. A particle is classified as \emph{stuck} if \(z_{\mathrm{rel}} < 1.1 D_n\) and \(v_x < v_{\mathrm{thresh}} = 0.05~\mathrm{m/s}\). Time-dependent fractions of stuck and suspended particles, duration in suspension and at rest, mean streamwise velocities, and bulk mass flux are computed. Particles are further categorized as \emph{ever moving} or \emph{ever stuck} depending on whether these conditions are met at least once. Scatter plots of rest vs suspension times and histograms of mobile vs immobile particle diameters summarize selective transport and sorting processes. Time series of mean relative height, stuck fraction, mean moving velocity, and bulk flux illustrate sediment dynamics at the individual and population level.
\begin{figure}[ht!]
	\centering   
	\includegraphics[width=1\linewidth]{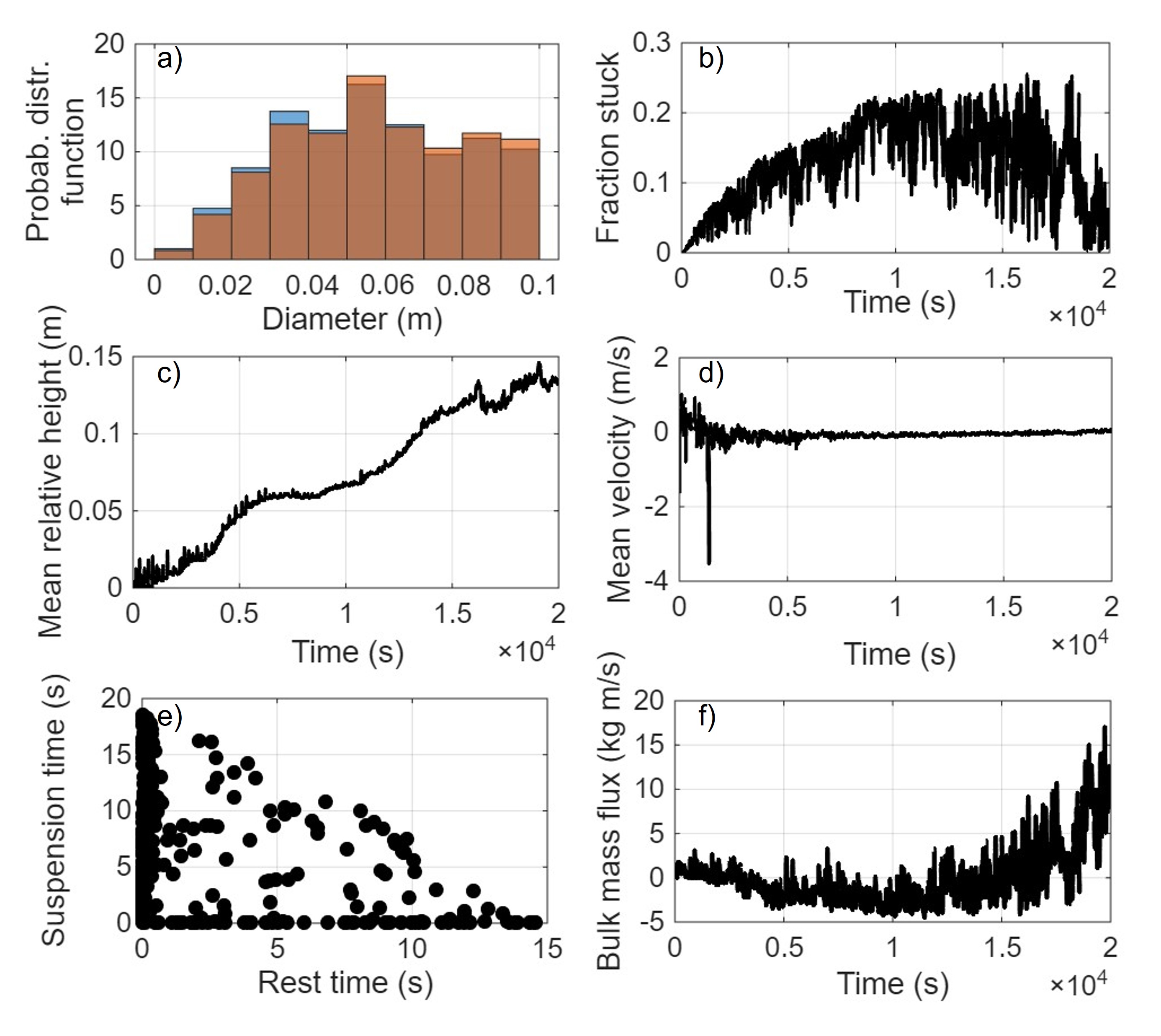}
	\caption{Sediment transport metrics from the numerical test case. (a) diameter distributions of \emph{ever moving} and \emph{ever stuck} particles (blue bins: always mobile, orange bins: never mobilized, brown: intermittent). (b) fraction of particles classified as stuck. The figure highlights intermittent transport, coexistence of mobile and immobile phases, and size-selective mobility. (c) mean relative particle height $\langle z_{\mathrm{rel}} \rangle(t)$. (d) mean streamwise vertical velocity of moving particles. (e) total rest vs suspension time for individual particles.  (f) bulk mass flux.}
	\label{fig:transport_metrics}
\end{figure}

Figure~\ref{fig:transport_metrics} summarizes sediment transport metrics. Diameter distributions, panel a, are used to characterize particle mobility as a function of grain size: blue bins represent grains that are consistently mobile, orange bins correspond to grains that remain immobile, and brown bins indicate particles that alternate between motion and rest. The partial overlap between these distributions highlights transport intermittency, as some grains transition between mobile and immobile states. Smaller grains are preferentially entrained, whereas larger grains tend to remain immobile, consistent with hiding–exposure effects and size-dependent entrainment thresholds observed in natural gravel-bed rivers \citep{Ancey2015}. The fraction of \emph{stuck} particles, panel b, is anti-correlated with $\langle z_{\mathrm{rel}} \rangle$, confirming that widespread vertical motion coincides with grain mobilization. The mean relative particle height $\langle z_{\mathrm{rel}} \rangle (t)$, panel c, provides a measure of sediment activity, as higher values indicate that particles are more frequently lifted above the bed and transported in partial suspension. 

Mean streamwise velocity, panel d, of moving particles captures the effective transport speed, responding to hydrodynamic forcing and inter-particle interactions. The bulk mass flux $\sum_i m_i v_{x,i}$, panel f, quantifies the instantaneous sediment transport rate, representing the total mass of particles crossing a given section per unit time. Peaks in this quantity correspond to episodes of enhanced transport activity driven by increased flow forcing or collective particle motion.

The numerical results demonstrate that the model reproduces key statistical features of bedload transport, including intermittency, collective mobilization, size-selective transport, and fluctuating bulk sediment flux. 

\section{Comparison with Field Seismic Data}
\subsection{Experimental site and data}
\begin{figure}[ht!]
	\centering
	\includegraphics[width=0.7\linewidth]{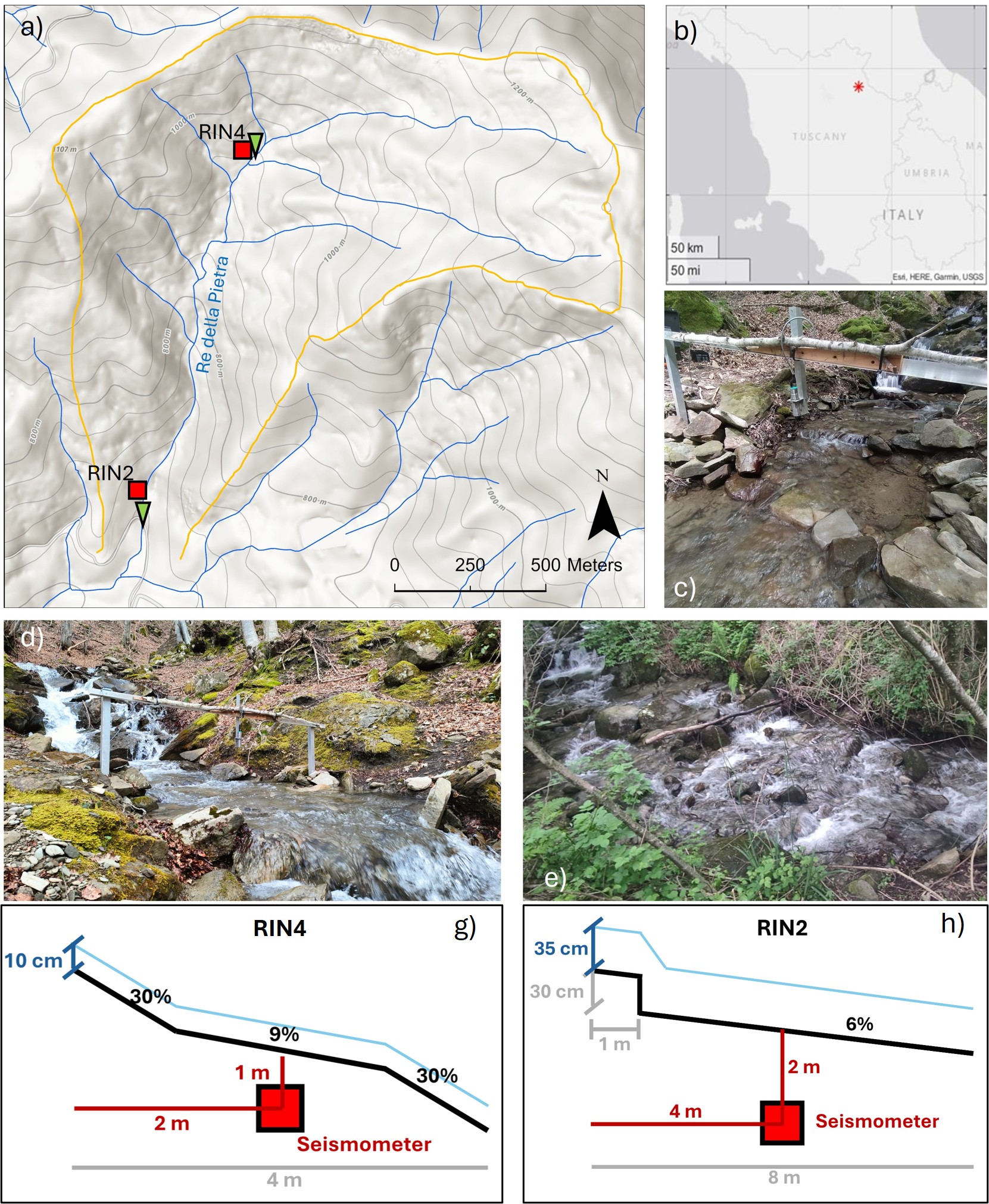}
	\caption{Map of the Re della Pietra catchment, showing the locations of the RIN2 and RIN4 seismic stations (red squares) and the flow altimeters (green triangles) (a). Panel showing the geographic location of the study site (red star) (b). Photographs of the Re della Pietra torrent at the RIN4 (c, d) and RIN2 (e) sections. The ultrasonic altimeter deployed at RIN4 is clearly visible in panels (c) and (d). Simplified sketches of the bed geometry at the RIN4 (g) and RIN2 (h) sections.}
	\label{mappa}
\end{figure}

To test our numerical model, we compare simulation results with seismic observations acquired during a flood event occurred on 30 May 2024 in the Re della Pietra torrent, a natural gravel-bed, step-pool mountain torrent located in the Rincine Forest, in the Tuscan Apennines (Figure \ref{mappa}). The event was recorded by two seismic stations (RIN2 and RIN4) deployed at two different sections of the torrent (Figure \ref{mappa}).
At both stations, ground velocity was recorded using a Lennartz LE-3D/5s triaxial seismometer, installed buried in the ground, and digitized at 200 Hz (consistent with the sampling rate used in the numerical simulations).

RIN4 (43.888°N, 11.627°E) was the upstream monitoring station, deployed at an elevation of 915 m a.s.l. near the catchment ridge (Figure \ref{mappa}). The seismometer was buried in the ground approximately 1 m from a pool of the torrent. The pool is about 4 m long and 1.5 m wide (Figure \ref{mappa}c, d, g) and is characterized by a bed composed of poorly sorted coarse material, ranging from fine gravel ($<$1 cm) to stones and boulders with diameters of several tens of centimeters (Figure \ref{mappa}c). The pool has an average bed slope of  9\% and is situated between two very steep torrent reaches (steps), where the channel inclination is about 30\% (Figure \ref{mappa}d, g). This configuration enhances flow energy and sediment motion at both the pool inlet and outlet. At this site, flow depth was continuously measured at 200 Hz using an ultrasonic distance sensor mounted above the water surface and positioned approximately at the center of the pool (Figure \ref{mappa}c–d).

RIN2 (43.878°N, 11.622°E) was deployed approximately 1200 m downstream of RIN4, at an elevation of 647 m a.s.l. (Figure \ref{mappa}a). At this site, the seismometer was buried in the bank, about 3 m from the torrent. Here, the Re della Pietra is a more developed mountain stream (Figure \ref{mappa}e) and exhibits a more regular channel geometry than the upstream reach (RIN4), with longer pools ($\sim$6–10 m) separated by short ($<$1 m) steps. In front of the seismometer, the channel bed consists of poorly sorted sediments ranging from gravel to stones (Figure \ref{mappa}e) and has an average slope of approximately 6\%. A step approximately 30 cm high is located at the upstream end of the monitored reach (Figure \ref{mappa}h). Near RIN2, flow depth is continuously recorded by a piezometer installed $\sim$30 m downstream of the seismometer (Figure \ref{mappa}a), with a 10-minute temporal resolution.\\

At both RIN2 and RIN4 stations, the recorded flood event was characterized by a markedly asymmetric hydrograph (Figure \ref{RIN4}a and \ref{RIN2}a) and produced an emergent seismic signal with the distinctive "cigar-shaped" envelope commonly associated with floods and other mass movements, such as debris flows and landslides \citep{Belli2022,Zobin2009} (Figure \ref{RIN4}b and \ref{RIN2}b). 
Despite these similarities, the event exhibits substantial differences in timing, duration, and flow depth between the two stations, reflecting surge propagation and contribution from tributary-torrents along the catchment. At the upstream station (RIN4), the flood peaked around 19:30 (UTC) and lasted approximately 6 hours, with flow depth rapidly rising from $\sim$4 to 12 cm before gradually receding (Figure \ref{RIN4}a). 
In contrast, at the downstream station (RIN2), the event persisted for more than 10 hours, with the water level increasing from $\sim$26 to 36 cm and subsequently slowly declining (Figure \ref{RIN2}a). The flood peak at RIN2 was recorded around 20:00 (UTC), half an hour later than that observed at RIN4. Considering the $\sim$1.2 km length of the channel reach between the two sites, this delay results in an estimated flood-wave velocity of $\sim$0.7 m/s.
The two seismic signals also exhibit different maximum amplitudes (around $7\times10^{-6}$ m/s at RIN4 and around $2\times10^{-6}$ at RIN2), also reflecting the different source-to-receiver distances (approximately 1 m for RIN4 and 3 m for RIN2). In addition, the signal recorded at RIN4 displays some sharp, higher-amplitude peaks during the phase of maximum flow depth, which are less pronounced in the signal recorded at RIN2.\\

\subsection{Seismic data analysis}
\begin{figure}[ht!]
	\centering
	\includegraphics[width=0.9\linewidth]{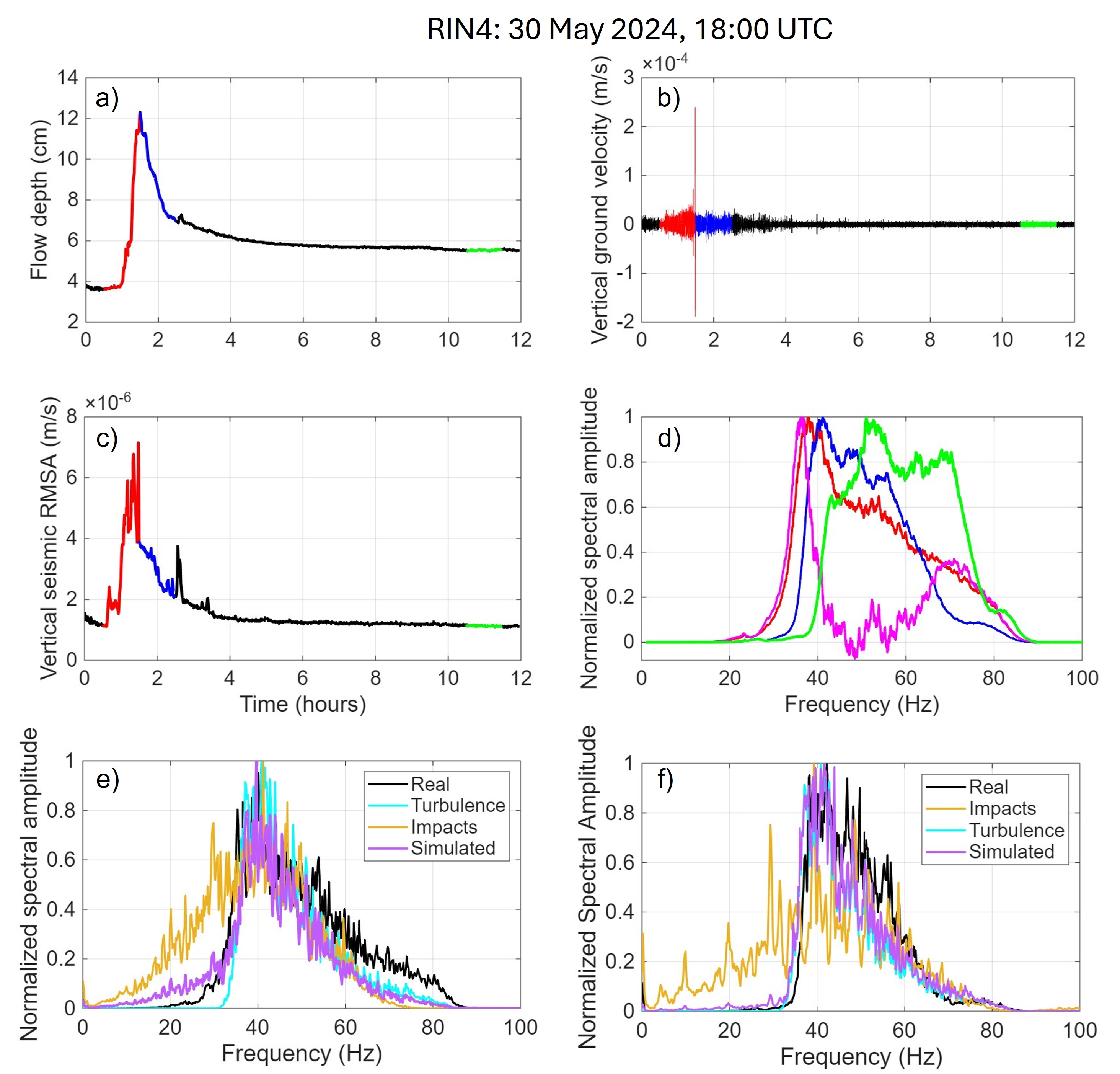}\\
	\caption{Flow depth (a) and vertical seismic signal (c) recorded by RIN4 seismic station during the flood event of 30 May 2024 in the Re della Pietra torrent. Colors indicate three different flow phases: rising limb (red), falling limb (blue), and final phase (green). Panel (c) displays the RMSA envelope of the vertical seismic signal. Power spectral density (PSD) spectra for the three flow phases are shown in panels (d); the magenta curve represents the difference between the rising and falling limb spectra. A consistent color coding is adopted across all plots: red, blue and green curves always correspond to the rising, falling and final flow phases, respectively. Panels (e–f) show the simulated seismic signals and their comparison with the observations at RIN4 for the rising (e) and falling (f) limbs of the flood. The best agreement with the observed spectra is obtained by assigning approximately 80\% of the total seismic energy to turbulent water forcing and 20\% to particle impacts. Simulation parameters are reported in Table 2.}
	\label{RIN4}
\end{figure}

\begin{figure}[ht!]
	\centering
	\includegraphics[width=0.9\linewidth]{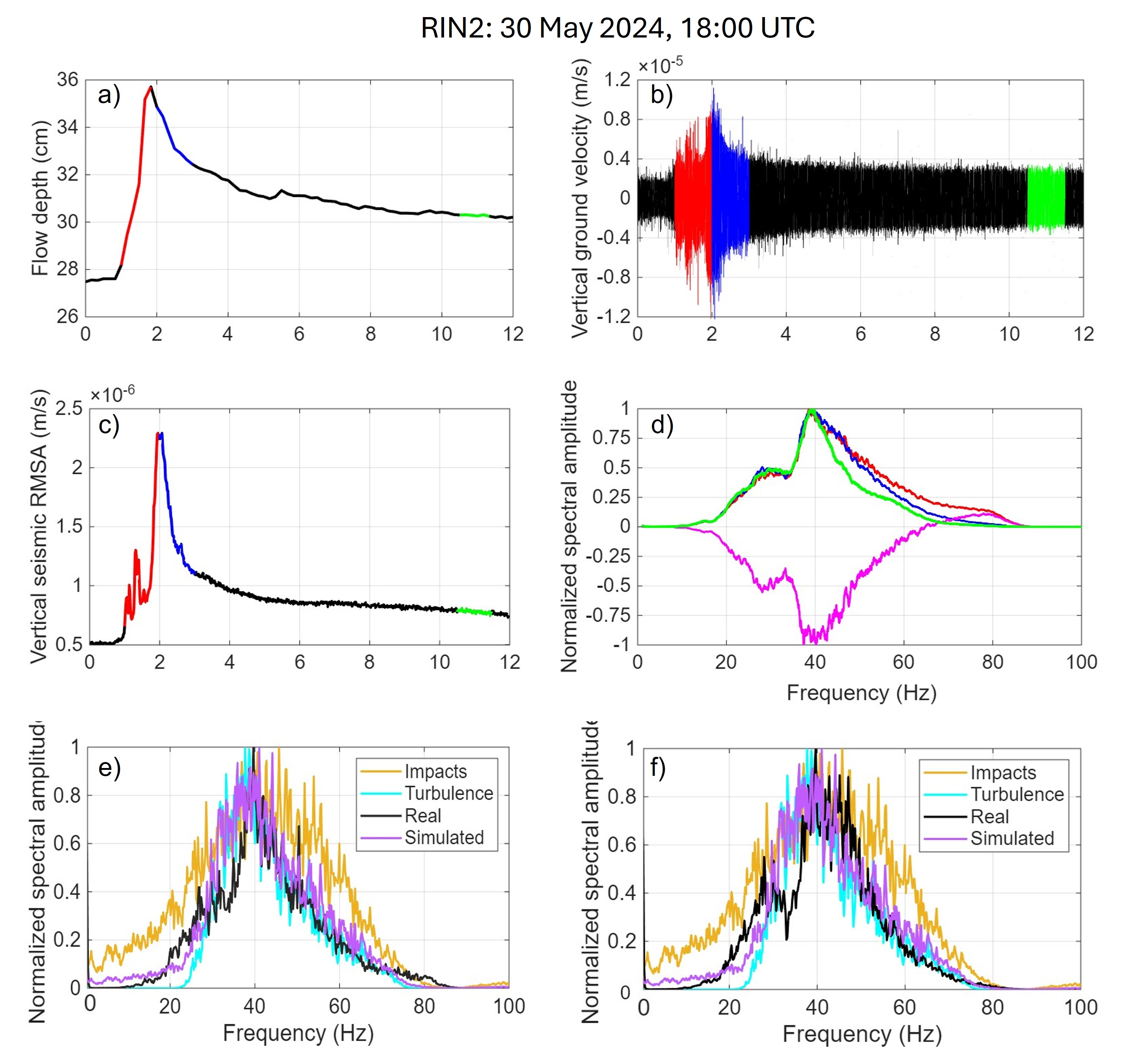}\\
	\caption{Flow depth (a) and vertical seismic signal (c) recorded by RIN2 seismic station and the nearby water-level sensor during the flood event of 30 May 2024 in the Re della Pietra torrent. Colors indicate three different flow phases: rising limb (red), falling limb (blue), and final phase (green). Panel (c) displays the RMSA envelope of the vertical seismic signal. Power spectral density (PSD) spectra for the three flow phases are shown in panels (d); the magenta curve represents the difference between the rising and falling limb spectra. A consistent color coding is adopted across all plots: red, blue and green curves always correspond to the rising, falling and final flow phases, respectively. Panels (e–f) present the simulated seismic signals and their comparison with the observations at RIN2 for the rising (e) and the falling (f) limbs of the flood. The best agreement with the recorded spectra is obtained by assigning about 75\% of the total seismic energy to water-induced turbulence and 25\% to particle impacts in both cases. Simulation parameters are summarized in Table 2.}
	\label{RIN2}
\end{figure}

To analyze seismic radiation at different stages of the event, we distinguish three 1-hour-long flood phases at both RIN4 and RIN2: the rising limb (red segments in Figures \ref{RIN4} and \ref{RIN2}), the falling limb (blue), and the final phase (green). These phases are expected to correspond to different water–sediment flow conditions, reflecting, respectively, sediment motion initiation associated with rising water levels and increasing bedload transport (red), decreasing flow energy and sediment transport (blue), and a return to near-normal flow conditions with negligible solid-particle motion (green).

For each flood phase, the recorded seismic signals are analyzed in both time and frequency domains. The root mean squared amplitude (RMSA) envelope of the seismic signal is computed for data in the 0.5–95 Hz range using 1-minute moving windows (Figure \ref{RIN4}c and \ref{RIN2}c). 
For the signal recorded at RIN4, despite the two phases being characterized by the same maximum flow depth, the rising limb of the flood produced higher seismic amplitudes than the falling limb (Figure \ref{RIN4}a, c). This difference is likely related to the higher energy of the seismic component generated by particle collisions at the onset of the flood compared to the recession phase. Indeed, during the initial stage of a flood, a larger amount of solid particles is generally available for transport along the riverbed; these particles are progressively entrained and depleted by the flow, becoming less abundant during the decreasing phase \citep{Belli2026}. Consequently, seismic radiation is stronger during the rising limb of the flood than during the falling limb. Instead, for the seismic component produced by water flow, no differences are expected for the same water level in the two flood phases. A similar discrepancy in the produced seismic amplitudes is not observed at RIN2, where the seismic signal shows nearly identical amplitudes during both flood phases (Figure \ref{RIN2}a, c), providing no evidence for preferential sediment transport during the rising limb of the flood at this torrent section.
This difference in sediment‐transport patterns during the flood at the two channel sections is likely related to variations in bed topography along the course of the Re della Pietra torrent. At RIN4, the torrent is steeper, narrower, and closer to the basin ridge (Figure \ref{mappa}c, d, g). These conditions favor the input of solid debris into the channel, enhance sediment transport, and promote progressive particle depletion during the initial phase of the flood. In contrast, at RIN2 the Re della Pietra is a more developed mountain torrent, characterized by higher discharge and a gentler bed slope. This setting likely results in a reduced input of new solid debris and a more uniform sediment transport throughout the flood.

We then compute the power spectral density (PSD) of the seismic signals recorded at both stations for each flood phase (Figure \ref{RIN4}d and \ref{RIN2}d). To facilitate comparison among phases characterized by different seismic amplitudes, all PSD spectra are normalized. Results reveal clear differences in seismic spectral radiation among the various flood phases, as well as between the two recording sites.\\
\noindent At RIN4, seismic energy is concentrated between $\sim$ 25 and 85 Hz during the rising limb of the flood and between 35 and 85 Hz during the falling limb, with a peak around 40 Hz in both phases (Figure \ref{RIN4}d). In the final phase of the event, seismic radiation is mainly observed between $\sim$ 40 and 80 Hz, with a peak around 50 Hz.
At RIN2, seismic energy is distributed between $\sim$ 15 and 85 Hz during the rising limb, between $\sim$ 15 and 80 Hz during the falling limb, and between $\sim$ 15 and 70 Hz during the final phase of the flood. In all phases, seismic radiation peaks at approximately 40 Hz.\\
\noindent To emphasize differences in seismic radiation between the rising and falling limbs of the flood at both seismic stations, we compute the difference between the spectra produced during the two phases prior to normalization (magenta normalized curves in Figures \ref{RIN4}d and \ref{RIN2}d). Significant discrepancy is observed in the spectral radiation recorded at RIN4. 
In particular, despite comparable flow depths, the rising limb exhibits higher seismic energy in the 20–40 Hz band and, to a lesser extent, between 65 and 80 Hz compared to the falling limb. 
In contrast, at RIN2 the falling limb is enriched in the 20–60 Hz frequency band, with a negative peak of the magenta curve at 40 Hz. This enrichment probably does not result from different source mechanisms during the two phases, but most likely reflects the higher average seismic energy recorded during the falling limb compared to the rising limb, leading to a higher-amplitude spectrum during the falling phase. For the rising limb, only a slight enrichment is observed between 60 and 85 Hz. \\
\noindent Since the RMSA analysis suggests preferential sediment transport during the rising limb of the flood at RIN4, the spectral differences observed between this phase and the falling limb (Figure \ref{RIN4}d) likely reflect the spectral signature of enhanced bedload transport during the rising limb. According to this hypothesis, the collision-related seismic source appears to dominate the seismic radiation at RIN4 in the 20-40 Hz and, to a lesser extent, 65-80 Hz frequency bands, whereas the water-flow-related source seems to prevail in the 40-65 Hz range (Figure \ref{RIN4}d). 

A secondary peak is visible at low frequencies in (Figure \ref{RIN4}d), possibly associated with particle rolling along the bed or hydrodynamic vortex shedding, although its contribution is minor compared to the dominant sources.

\subsection{Numerical modeling of recorded seismic data}

To reproduce the spectral characteristics of the recorded seismic signals, synthetic ground velocity was constructed by combining particle-impact and turbulent water-forcing contributions, computed as described in Section \ref{Methods}.  For the comparison with the simulations, two representative time windows were selected from the seismic record: one during the rising limb of the flood and one during the decreasing phase of the event. In each case, a two-minute segment of the signal was extracted and used to compute the corresponding power spectral density.
Although the numerical framework also allows simulation of rolling contacts and vortex-shedding forcing, these mechanisms were not included in the final synthetic signal because the observed PSD does not show clear spectral signatures associated with them. Including these sources degraded the agreement with the observations.
Simulations were conducted using the actual geometries of the RIN4 and RIN2 channel sections (schematized in Figure \ref{mappa}g-h) and input hydraulic parameters (Table \ref{tab_params}) consistent with recorded flow depth and grain size sampled in the study site. In particular, flow velocity was estimated on the basis of recorded flow depth and channel geometry, using the Manning equation \citep{Manning1891,Chaudhry2008}.
Based on the RMSA analysis of seismic data, which indicates preferential bedload transport during the rising limb of the flood at RIN4, we set $N_p=2$ for the rising limb and $N_p=0.4$ for the falling limb (Table \ref{tab_params}). In contrast, the RIN2 seismic data do not suggest enhanced transport during the initial phase of the flood; therefore, a constant value of $N_p=2$ was used for both the rising and falling limbs (Table \ref{tab_params}). In the simulations, the turbulent signal is generated within a prescribed frequency band that differs between the two sites: for RIN4, the spectral range is set to $30–90$ Hz, whereas for RIN2 it is extended to $20–90$ Hz. This choice reflects the greater flow depth recorded at RIN2, which promotes the development of larger turbulence structures and a consequent shift of the radiated energy toward lower frequencies \citep{Gimbert2014,Coco2021,Belli2022}.

\begin{table}[ht!]
	\centering
	\caption{Summary of the main parameters used in the numerical simulation for RIN4 and RIN2 seismic stations.}
	\label{tab:RIN4}
	\begin{tabular}{lcccccc}
		\hline
		RIN4 & $U_0$ [m/s] & $H$ [cm] & $N_p$  & $D_{med}$ [cm] & $T$ [s] \\
		\hline
		rising limb & 1.3 & 10 & 2  & 5 & 120 \\
		falling limb & 1.3 & 10 & 0.4  & 5 & 120 \\
		\hline%
		RIN2 & $U_0$ [m/s] & $H$ [cm] & $N_p$  & $D_{med}$ [cm] & $T$ [s] \\
		\hline
		rising and falling limb & 1.7 & 35 & 2  & 5 & 180 \\
	\end{tabular}
	\label{tab_params}
\end{table}

The relative contributions of particle impacts and turbulent forcing were adjusted to reproduce the observed PSD of seismic signals recorded at stations RIN4 and RIN2. The best agreement for the upstream station RIN4 is obtained when turbulent forcing contributes approximately 80\% of the total seismic energy, with particle impacts accounting for the remaining 20\%, consistently during both the rising and decreasing phases of the event. At the downstream station RIN2, a slightly larger contribution from particle impacts is required, with approximately 75\% of the energy associated with turbulence and 25\% with particle impacts.

\section{Discussion}

Comparison between simulated and observed spectra demonstrates that the model captures the primary characteristics of the seismic signals recorded during both the rising limb and the final flood phase at both stations. Specifically, the spectral amplitude distribution, overall PSD shape, and peak frequency are well-reproduced when the synthetic signal combines turbulent forcing and particle-impact contributions. The main discrepancies occur at frequencies above 70 Hz during the rising limb, where simulated spectra appear depleted relative to the observed data. These differences may stem from either the model's inability to fully reproduce high-frequency seismicity or from additional environmental sources active in the field, such as rainfall \citep{Bakker2022}, which are not currently incorporated into the model.

The primary characteristics of the recorded spectra were successfully captured in the simulations.
The presented numerical framework also makes it possible to evaluate the relative contributions of several potential processes, including particle impacts, rolling contacts, turbulent flow forcing, and vortex-shedding effects.
This allows for an investigation of the seismic source mechanisms controlling fluvial seismic noise during flood events.

Our results indicate that the recorded signals are primarily controlled by two dominant mechanisms: turbulent hydrodynamic forcing and impulsive forces produced by grain–bed impacts. This finding is consistent with previous theoretical models and experimental observations of fluvial seismicity \citep{Burtin2008,Tsai2012,Gimbert2014,Schmandt2013,Roth2017}. 
The relative importance of the two sources, however, varies spatially. At station RIN4, the spectra are best reproduced when turbulent forcing accounts for approximately 80\% of the total energy, with particle impacts contributing the remaining 20\% across both flood phases. At the downstream station RIN2, the particle impact contribution increases slightly, reaching 25\% during the rising limb and 30\% during the decreasing phase. This shift likely reflects spatial variations in sediment transport conditions, channel geometry and flow parameters along the torrent, with enhanced particle activity at the downstream site.\\
Results therefore suggest that turbulent flow represents the dominant seismic source during high-discharge conditions in mountain torrents, in agreement with findings by \citet{Roth2017}.
On the other hand, bedload transport provides significant contribution in the high-frequency (greater than 10 Hz) band through grain impacts. The balance between these mechanisms is expected to depend on local hydraulics, sediment supply, and channel morphology.

Furthermore, our model enables a preliminary quantification of the seismic energy produced by bedload transport versus water flow, providing a valuable tool for discriminating between these two source mechanisms in natural streams. This paves the way for further advancements in the seismic-based quantification of bedload transport. 

\noindent While the model can simulate additional processes such as rolling contacts and vortex shedding, these components do not produce identifiable spectral signatures in the observed data. In fact, their inclusion introduces spectral features not present in the field observations, thereby reducing the overall agreement. Consequently, the final synthetic signals were constructed using only turbulent water forcing and particle impacts.

Despite the general agreement with observations, some limitations remain in modeling river seismicity. 

The weights used to combine turbulent forcing and particle-impact contributions are treated as effective parameters, accounting for uncertainties in source efficiency, ground coupling, and wave propagation, rather than representing strict physical partitions between the underlying processes. Moreover our simplified representation of particle dynamics in terms of ideal spheres may not fully capture the complexity of natural transport, such as heterogeneous grain sizes or intermittent movement.

\section{Conclusions}

This study presents a numerical framework to investigate the physical processes responsible for seismic noise generated by sediment transport during flood events in natural rivers and torrents. The model simulates particle trajectories, particle–bed interactions, and turbulent flow forcing, allowing the different source mechanisms of fluvial seismic signals to be evaluated within a unified framework.

The comparison between synthetic and observed seismic spectra radiated by a small mountain torrent shows that the recorded signal can be interpreted as the combined effect of turbulent hydrodynamic forcing and impulsive particle impacts with the river bed. Turbulent flow forcing provides the dominant contribution to the seismic energy during high-discharge conditions, while particle impacts contribute additional high-frequency energy associated with bedload transport. Although our results indicate that particle impacts and turbulent hydrodynamic forcing dominate the seismic signal in our study area, it is important to contextualize the role of the additional modeled processes. As previously mentioned, the inclusion of pure rolling contacts and vortex shedding slightly worsened the fit with the experimental data, leading us to consider them as secondary sources for this specific event. Nevertheless, the relative contribution of these sources could vary significantly under different geomorphological and hydrodynamic conditions. For example, in rivers characterized by gentler slopes, different bed roughness, or distinct grain-size distributions, prolonged rolling contacts and the specific frequencies generated by vortex shedding might become non-negligible, thus playing a more prominent role in the overall seismic budget.

The proposed approach provides a physically based link between sediment transport dynamics and seismic observations. This represents an important step toward the quantitative interpretation of fluvial seismic signals and their use as a proxy for sediment transport processes.
Future developments of the model could incorporate more realistic river-bed topography, enhanced treatment of particle-particle interactions, and the analysis of longer seismic time series to better constrain sediment transport dynamics during natural floods. Future applications could be extended to higher-magnitude events, such as major floods in large rivers, or debris flows, where sediment transport is extreme compared to rivers and has been shown to dominate seismic radiation.

\section{Acknowledgment}
The work was developed under the auspices of GNFM (INdAM) and received financial support by projects TRANSFORM (CUP B55F21007810001), PRIN-22 (CUP B53D23009510006).

\bibliographystyle{apalike}
\bibliography{references}

\end{document}